\documentclass{ws-procs9x6}            

\usepackage{graphicx}
\usepackage{subfigure}
\usepackage{mathrsfs}
\usepackage{multirow}
\usepackage{lscape}

\begin{document}
\title{Computation of the Binding Energies in\\ the Inverse Problem Framework}

\author{S. Cht. ~Mavrodiev}

\address{The Institute for Nuclear Research and Nuclear Energy, BAS,\\
Sofia, 1784, Bulgaria\\ 
E-mail: schtmavr@yahoo.com\\
}

\author{M.A.~Deliyergiyev}

\address{Department of High Energy Nuclear Physics, \\
Institute of Modern Physics, CAS, Lanzhou, 730000, China\\
$^*$E-mail: deliyergiyev@impcas.ac.cn\\
}

\begin{abstract}
We formalized the nuclear mass problem in the inverse problem framework. This approach allows us to infer the underlying model parameters from experimental observation, rather than to predict the observations from the model parameters. The inverse problem was formulated for the numericaly generalized the semi-empirical mass formula of Bethe and von Weizs\"{a}cker. It was solved in step by step way based on the AME2012 nuclear database. 

The solution of the overdetermined system of nonlinear equations has been obtained with the help of the Aleksandrov's auto-regularization method of Gauss-Newton type for ill-posed problems. 
In the obtained generalized model the corrections to the binding energy depend on nine proton (2, 8, 14, 20, 28, 50, 82, 108, 124) and ten neutron (2, 8, 14, 20, 28, 50, 82, 124, 152, 202) magic numbers as well on the asymptotic boundaries of their influence. These results help us to evaluate the borders of the nuclear landscape and show their limit. The efficiency of the applied approach was checked by comparing relevant results with the results obtained independently. 
\end{abstract}

\keywords{Bethe-Weizs\"{a}cker Mass Formula;  Magic Numbers; Binding energy; Inverse Problem.}

\bodymatter

\section{Introduction}
The main goal of our studies was to determine how well the existing data, and only data, determines the mapping from the proton and neutron numbers to the mass of the nuclear ground state. Another is to find presumed regularities by analysis of observed nuclei masses \cite{AleksandrovGadjokov1971}. In addition is to provide reliable predictive model that can be used to forecast mass values away from the valley of stability. 

The aforementioned goals stimulated us to try to clarify the features and to find hidden regularities of the well known semi-empirical mass formula of Bethe-Weizs\"{a}ker (BW), based exclusively on experimental data. A set of experimental nuclear masses from AME2012, the most recent evaluation database, that was published in December 2012 in \cite{Ami2012_ChinPhysC}, constitutes the raw material for this work.  Only measured nuclei are included into our consideration. The masses extrapolated from systematics and marked with the symbol $\#$ in the error column are not taken into account here. Therefore we use only 2564 experimental nuclear masses, including the Hydrogen atom, to provide a deep understanding of the mutual influence of terms in the semi-empirical BW mass formula. 

In present work we demonstrate the applicability of the inverse problem (IP) approach for solution of such nuclear physics problems. First we formalize the nuclear mass problem in the framework of the IP. Second we propose the generalized form of the BW mass formula, which helps us to discover the latent regularities in the nuclear masses. 
Afterwards we provide a solution of the formalized IP that has been obtained with the help of the Alexandrov dynamic auto-regularization method of the Gauss-Newton type for ill-posed problems (REGN-Dubna) \cite{Alexandrov:1973, Aleksandrov197146, Alexandrov66:1977}, which is a constructive development of the Tikhonov regularization method \cite{Tikhonov:1963, Tikhonov:1983} for ill-posed problems. 

The formalism of the applied approach is given in Sec.~\ref{Theory_and_Method}. 
The numerical generalization of the BW mass formula is described in Sec.~\ref{Parametrization_of_the_BWmass_formulae}. 
The main conclusions of the paper are drawn in Sec.~\ref{Results}, which also includes a discussion of the principal results, the resulting rms deviations.

\section{Theory and Method}
\label{Theory_and_Method}


IPs are based on the comparison of the theoretical and experimental data by solving the system of the nonlinear operator equations on the field of real numbers $\bm{R}$  of following type: 
\begin{equation}
F(x) = y,
\label{eq:fxy_VectorForm}
\end{equation}
where  $x=(x_{1},x_{2},\dots,x_{n})^{T} \in \bm{R}^{n}$, $y=(y_{1},y_{2},\dots,y_{m})^{T} \in \bm{R}^{m}$. 
$\bm{R}^{n}$ and $\bm{R}^{m}$ are the $n$ and $m$ dimensional real coordinate space, that corresponds to the $x$ and $y$ respectively.

The Eq.\eqref{eq:fxy_VectorForm} typically involve the estimation of certain quantities based on indirect measurements of these quantities. The estimation process is often ill-posed in the sense that noise in the data may give rise to significant errors in the estimate. In other words, the problem Eq.\eqref{eq:fxy_VectorForm} is ill-posed \cite{TikhonovArsenin:1977} if its solution does not depend continuously on the right hand side $y$, which is often obtained by measurement and hence contains errors, $\Arrowvert  y^{\delta}-y \Arrowvert ~\leq\delta$,  where $y^{\delta}$ is the measured perturbed data, $\delta$ is the experimental uncertainty. 
Operator $F$  is a forward modeling nonlinear operator, that transforms any model $x$ into the corresponding data $y$. The IP is formulated as the solution of the operator equation. Therefore, Eq.\ref{eq:fxy_VectorForm} connects the unknown parameters of the model with some given quantities (variables) describing the model, in our case atomic mass number, $A$, and proton mass number, $Z$. These quantities take the form of the so-called input data. 

Techniques known as regularization methods \cite{TikhonovArsenin:1977, Alexandrov:1973, Aleksandrov197146, Tikhonov:1983} have been developed to deal with this ill-posedness, to get stable approximations of solutions of Eq.\eqref{eq:fxy_VectorForm}. In the current work we apply the Alexandrov method\cite{Alexandrov:1973, Aleksandrov197146}, which we found an appropriate choice for our IP that will be formulated in the next section.

\section{Parameterization of the Bethe-Weizs\"{a}cker mass formula}
\label{Parametrization_of_the_BWmass_formulae}

A careful analysis of the previous attempts of calculation nucleus masses reveals that all models can reproduce experimental/empirical trends on the average 
\cite{
Moller1995185, 
Wapstra2003129, 
Chowdhury:2004jr, 
Kirson200829,
WDMyers:1966,
Myers1996141, 
Samanta:2004et, 
Rohlf:1994, 
NIX19651, 
Wapstra:1958,
PhysRev.89.1102, 
PhysRevC.33.2039,
PhysRevC.67.044316,
Royer20131}. 
In order to overcome this issue we consider a parameterized nonlinear dynamical system of equations for determining nuclei and atomic masses from the experimental bound-state energies, which can be written using matrix notation of Eq.\ref{eq:fxy_VectorForm} as: 
\begin{equation}
\begin{split}
&FE_{B,j}^{\rm{Th}}(A,Z,\{a_{i}\}) =E_{B,j}^{\rm{Expt}}(A,Z),\\
&FM_{a.m.,j}^{\rm{Th}}(A,Z,\{a_{i}\}) = M_{a.m.,j}^{\rm{Expt}}(A,Z),\\
&FM_{n.m.,j}^{\rm{Th}}(A,Z,\{a_{i}\}) = M_{n.m.,j}^{\rm{Expt}}(A,Z),\\
&F\Delta m_{j}^{\rm{Th}}(A,Z,\{a_{i}\}) = \Delta m_{j}^{\rm{Expt}}(A,Z),
\label{eq:BE_j}
\end{split}
\end{equation}
where $F$ is the rectangular $d\times m$ $\{d=1,\dots,{\mathcal{N}}_{\rm{data}}; m=1,\dots,{\mathcal{N}}_{\rm{param}}\}$ Jacobian matrix composed of the Frechet derivatives with respect to the model parameters. Each system of Eqs.\eqref{eq:BE_j} contains 2564 equations, which correspond to the number of the experimental data-points. Therefore, these systems are overdetermined because the number of considered equations exceeds the number of parameters used in the fit. The right hand side of these equations represented by the vector of the observed experimental data, $j$ is the component of this vector, also it indicates the nonlinear equation in the system.

The solution of the overdetermined system of Eqs.\eqref{eq:BE_j} for the binding energy and its model is given by the real values of the parameters $a_{i}$ $(i=1,\dots, {\mathcal{N}}_{\rm{param}})$.

In order to proceed further one have to choose the initial model for description of the binding energy. The BW mass formula was chosen for this role, since it provides the baseline fit for all the rest models 
\cite{
Moller1995185, 
Wapstra2003129, 
Chowdhury:2004jr, 
Kirson200829,
WDMyers:1966,
Myers1996141, 
Samanta:2004et, 
Rohlf:1994, 
NIX19651, 
Wapstra:1958,
PhysRev.89.1102, 
PhysRevC.33.2039,
PhysRevC.67.044316,
Royer20131}. 
The generalization of the BW mass formula for the binding energy per nucleon in our approach has the following form:
\begin{equation}
\begin{split}
&E_{B}(A,Z,\{a_{i}\}) = \alpha_{vol}(A,Z,\{a_{i}\}_{1})-\alpha_{surf}(A,Z,\{a_{i}\}_{2}) \frac{1}{A^{p_{1}(A,Z, \{a_{i}\}_{6})}}\\
&-\alpha_{comb}(A,Z,\{a_{i}\}_{3})\frac{Z(Z-1)}{A^{p_{2}(A, Z, \{a_{i}\}_{7})}} -\alpha_{sym}(A,Z, \{a_{i}\}_{4} )\frac{(N-Z)^{2}}{A^{p_{3}(A,Z, \{a_{i}\}_{8} )}}\\
&+\alpha_{Wigner}(A,Z,\{a_{i}\}_{5})\frac{\delta(A,Z)}{A^{p_{4}(A, Z, \{a_{i}\}_{1}) }}+K_{MN}(A,Z,\{a_{i}\}_{10}).\\
\end{split}
\label{eq:BWparametrization_Wigner_CorMN}
\end{equation}
The term $\delta(A,Z)$ is equal to 1 for even~$N,Z$; -1, for odd $N,Z$ and 0, for odd $(Z+N)$. The detailed description of all the rest terms of Eq.\eqref{eq:BWparametrization_Wigner_CorMN} is beyond the focus of the current note, therefore for a curious reader we recommend this Ref.\cite{Mavrodiev:2016upv} for more detailed examination. However, note, that totally in our parametrization we use 17 linearly independent variables and 241 free parameters.

In order to solve all systems of Eqs.\eqref{eq:BE_j} we adopt the following procedure. First, we searched for the appropriate solution for the binding energy for the given $A$ and $Z$. The obtained solution was used as an approximation (`fake solution') to find the atomic masses, the solution for atomic masses was in turn used to find nuclear mass. Then, the fit is performed again, allowing the binding energy to vary for the given $A$ and $Z$. The resultant binding energy is then taken to be a new seed, and this procedure is repeated iteratively until convergence is reached for all systems. Indeed the system converge due to applied iterative process and due to theorems proved in Ref.\cite{Alexandrov:1973, Aleksandrov197146}.

\begin{figure}
\centering
\includegraphics[scale=0.18]{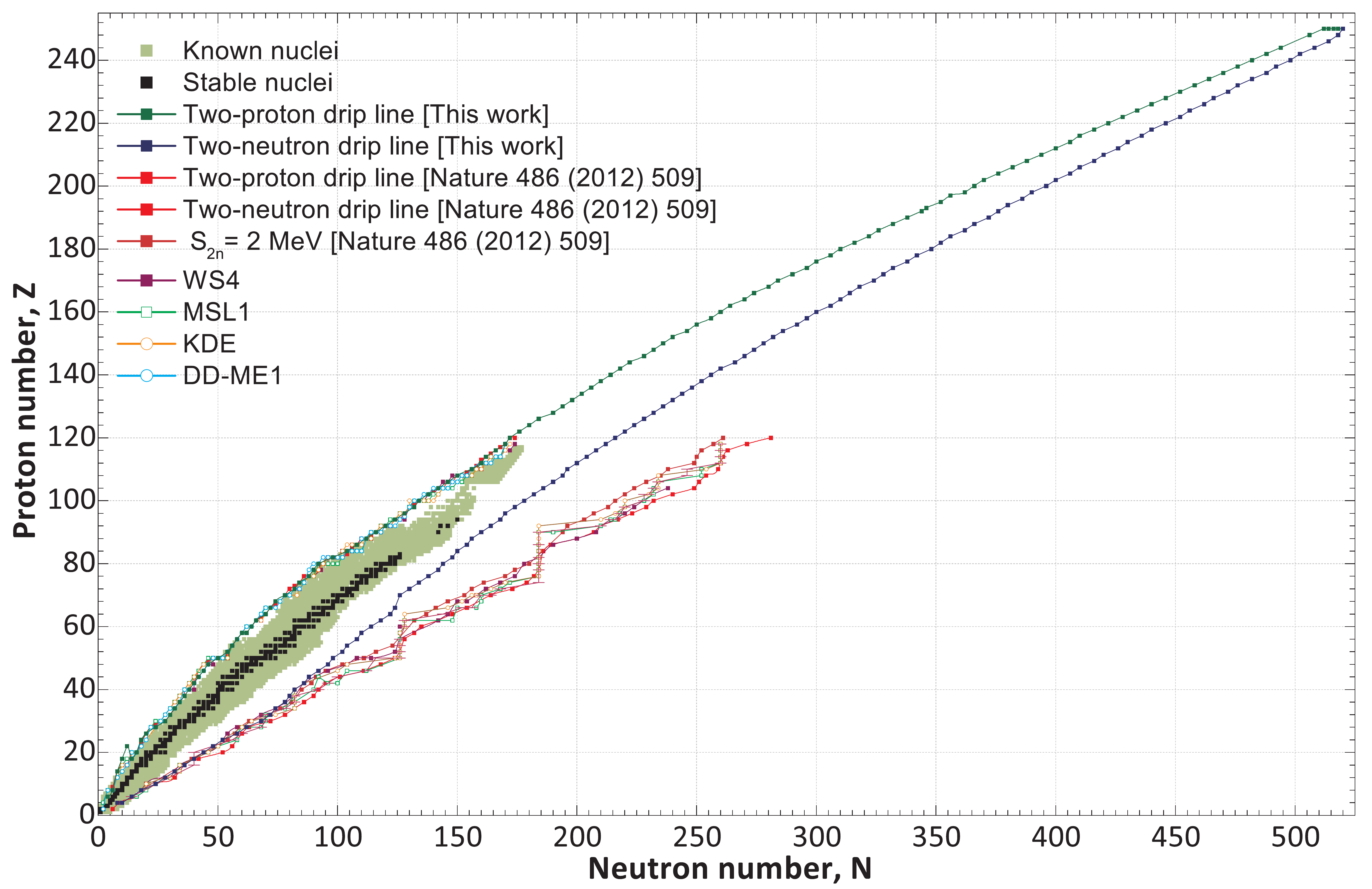}
\caption{
The chart of nuclei  known experimentally \cite{Ami2012_ChinPhysC} as a function of $Z$ and $N$. The stable (black squares) and radioactive (light green squares). Mean drip lines (red) are shown together with the $S_{2n}=2$ MeV line (brown).  The landscape of bound even-even nuclei as obtained from the microscopic density functional theory (DFT) calculations with two Skyrme interactions KDE \cite{PhysRevC.85.024305}, MSL1 \cite{Zhang2013234} and one relativistic interaction, DD-ME1 (cyan circles) \cite{PhysRevC.89.054320}. The prediction from Weizs\"{a}cker-Skyrme mass formula with WS4 \cite{Wang2014215} is also included for comparison (purple squares). The data is extracted from Ref. \cite{Erler_nature:2012, Wang:2014mra} and references therein. Predictions for the two-neutron (olive) and two-proton drip (royal) lines are also shown. 
}
\label{fig:DripLines_Predictions}
\end{figure}

\section{Results}
\label{Results}


The presented parameterization of the binding energy allows us to solve the formulated IP, with the help of the Aleksandrov's auto-regularized method. This solution provides us description of the 2564 nucleus masses and their binding energies starting from $^{2}_{1}$H with relative error $-1.1924\times10^{-6}$ and $3.2197\times10^{-4}$ for atomic mass and binding energy respectively. 

Mean absolute error, $\bar{\epsilon}_{abs}$, is $-0.0509$ and $7.8488\times10^{-4}$ for atomic mass and binding energy respectively. The maximum absolute deviation is less than 2.6 MeV for the atomic mass and less than 0.82 MeV for the binding energy. 

The mean modified error, $\bar{\epsilon}_{\rm{mod}}$ , of our solution is of 1.232(1\%) $\textendash$ 1.231(5\%) MeV, the $\sigma_{\rm{mod}}$ of 1.11(1\%) $\textendash$ 1.565(5\%) MeV (assuming that the theoretical uncertainty varies between 1\% and 5\%) for the atomic mass. While for the binding energy, using the same assumption for band of the theoretical uncertainties, we obtain $\bar{\epsilon}_{\rm{mod}}$, of fit of 0.044(1\%) $\textendash$ -0.0057(5\%) MeV, the $\sigma_{\rm{mod}}$ of 0.209(1\%) $\textendash$ 0.223(5\%) MeV,  which can be compared with the latest fits for all modern mass formulas \cite{PhysRevLett.16.197, Janecke1988265, Liran1976431, Nayak1999213, Koura200047, PhysRevC.52.R23}.



The obtained results and validation of our approach on the big set of the experimental data allows us to probe the outer boundary of the island of enhanced nuclear stability.  
The reshaping of the nuclear landscape, illustrated in Fig.\ref{fig:DripLines_Predictions}, is the results of such probe. 
In finding both the proton and neutron drip lines, two neutron ($S_{2n}$) and two proton ($S_{2p}$) separation energies were used.  The two-neutron and two proton drip lines are reached when $S_{2n}\approx 0$ and $S_{2p}\approx 0$, respectively. 

\section{Summary}
\label{Summary}

We found that results of the verification of the Bethe-Weizs\"{a}cker mass formula in the IP framework greatly improve the agreement between the experimental masses and the calculated ones and thus predicts the drip-lines more accurately than calculated earlier \cite{PhysRevC.65.037301, ADHIKARI:2004, BASU:2004} with the modified BW mass formula alone. This in turn allows us to make a prediction for the binding energy, nuclear and atomic mass, and mass excess of the recently discovery nuclei at the DGFRS separator \cite{Oganessian201562}, see Ref.\cite{Mavrodiev:2016upv}.
Moreover, this finding together with the asymptotic behavior, and accurate predictions of the binding energies of all known isotopes allows us to obtain quite precise predictions for the location of the proton and neutron drip lines, and claim the exact number of bound nuclei in the nuclear landscape, see Fig.\ref{fig:DripLines_Predictions}. We would like to comment that this result is not in a conflict with respect to the problem of the critical charge in QED, for interested reader we refer to \cite{Popov2001}, where one may find all necessary details. 

We are of course aware that the super-heavy isotopes in Fig.\ref{fig:DripLines_Predictions} cannot be produced at present and may be even at future facilities and that, so far, only theoretical studies may be carried out in such region of the nuclear chart.

The concept of ill-posed problems and the associated regularization theories seem to provide a satisfactory framework to solve nuclear physics problems. This new perspective can be used for testing the applications of the liquid model in different areas, for forecasting the mass values away from the valley of stability, for calculation of kinetic and total energy of nuclear proton, alpha, cluster decays and spontaneous fission, for preliminary research of island stability problem and the possibility of creating new super-heavy elements in the future.

\section*{Acknowledgments}
We are grateful to {\bf{Lubomir Aleksandrov}} for the provided REGN program. S. Cht. Mavrodiev would like to thank {\bf{Alexey Sissakian}} and {\bf{Lubomir Aleksandrov}} for many years of a very constructive collaboration and friendship. Unfortunately, they passed away, but they made enormous contributions to the REGN development and application of it for solving different IPs, especially for the discovering latent regularities. 

This work was partially supported by the Chinese Academy of Sciences President's International Fellowship Initiative under Grant No. 2016PM043.

\bibliographystyle{ws-procs9x6} 
\bibliography{numgen_BWmf_references}

\begin{thebibliography}{10}

\bibitem{AleksandrovGadjokov1971}
L.~Aleksandrov and V.~Gadjokov, {Analysis of Latent Regularities by means of
  Regularized Iteration Processes}, {\em {Journal of Analytical Chemistry}}
  {\bf 9}, 279  (1971), {[in Russian] Comm. JINR P5-5137 (Dubna, 1970)}.

\bibitem{Ami2012_ChinPhysC}
G.~Audi {\em et~al.}, {The AME2012 atomic mass evaluation}, {\em Chinese
  Physics C} {\bf 36}, p. 1287  (2012).

\bibitem{Alexandrov:1973}
L.~Aleksandrov, {The program REGN (Regularized Gauss-Newton iteration method)
  for solving nonlinear systems of equations}, {\em {USSR Comput. Math. and
  Math. Phys.}} {\bf 11}, 36  (1970), {[in Russian] Zh. Vychisl. Mat. Mat.
  Fiz., JINR Dubna Preprints: P5-7258, P5-7259 (Dubna, 1973)}.

\bibitem{Aleksandrov197146}
L.~Aleksandrov, {The Newton-Kantorovich regularized computing processes}, {\em
  {USSR Comput. Math. and Math. Phys.}} {\bf 11}, 46   (1971).

\bibitem{Alexandrov66:1977}
L.~Aleksandrov, {On Numerical Solution on Computer of the Nonlinear Noncorrect
  Problems}, {\em Comm. JINR P5-10366}   (1977).

\bibitem{Tikhonov:1963}
A.~N. Tikhonov, {About Regularization of the Ill-Posed Problems}, {\em Doklad
  Academy of Science USSR} {\bf 153}, 49  (1963), [in Russian].

\bibitem{Tikhonov:1983}
A.~N. Tikhonov {\em et~al.}, {\em {Regularizing Algorithms and A Priori
  Information}} ({Nauka}, Moscow, 1983), {[in Russian]}.

\bibitem{TikhonovArsenin:1977}
A.~N. Tikhonov and V.~Y. Arsenin, { Solutions of ill-posed problems, }{Scripta
  Series in Mathematics} ({V.H. Winston \& Sons}, {Washington, D.C.}, 1977) pp.
  xiii+258.
\newblock {}.

\bibitem{Moller1995185}
P.~M\"{o}ller {\em et~al.}, Nuclear ground-state masses and deformations, {\em
  Atom.Data Nucl.Data Tabl.} {\bf 59}, 185   (1995).

\bibitem{Wapstra2003129}
A.~Wapstra, G.~Audi and C.~Thibault, {The AME2003 atomic mass evaluation: (I).
  Evaluation of input data, adjustment procedures}, {\em Nucl. Phys. A} {\bf
  729}, 129   (2003).

\bibitem{Chowdhury:2004jr}
P.~R. Chowdhury and D.~N. Basu, {Nuclear matter properties with the
  re-evaluated coefficients of liquid drop model}, {\em Acta Phys. Polon.} {\bf
  B37}, 1833  (2006), \textcolor{blue}{[arXiv:nucl-th/0408013]}.

\bibitem{Kirson200829}
M.~W. Kirson, {Mutual influence of terms in a semi-empirical mass formula},
  {\em Nucl. Phys. A} {\bf 798}, 29   (2008).

\bibitem{WDMyers:1966}
W.~D. Myers and W.~J. Swiatecki, Nuclear masses and deformations, {\em Arkiv
  f\"{o}r Fysik} {\bf 36}, p. 343  (1966).

\bibitem{Myers1996141}
W.~Myers and W.~Swiatecki, {Nuclear properties according to the Thomas-Fermi
  model}, {\em Nucl. Phys. A} {\bf 601}, 141   (1996).

\bibitem{Samanta:2004et}
C.~Samanta and S.~Adhikari, {Shell effect in Pb isotopes near the proton drip
  line}, {\em Nucl. Phys.} {\bf A738}, 491  (2004),
  \textcolor{blue}{[arXiv:nucl-th/0402016]}.

\bibitem{Rohlf:1994}
J.~Rohlf, {\em { Modern Physics from alpha to $Z^{0}$}} ({John Wiley \& Sons},
  Canada, March 1994), {}.

\bibitem{NIX19651}
J.~R. Nix and W.~J. Swiatecki, Studies in the liquid-drop theory of nuclear
  fission, {\em Nucl. Phys.} {\bf 71}, 1   (1965).

\bibitem{Wapstra:1958}
A.~Wapstra, {Atomic Masses of Nuclides}, in {\em {External Properties of Atomic
  Nuclei / \"{A}ussere Eigenschaften der Atomkerne}\/},  ed. S.~Flügge,
  {Encyclopedia of Physics / Handbuch der Physik}, Vol.~8/38/1 (Springer Berlin
  Heidelberg, 1958) pp. 1--37.

\bibitem{PhysRev.89.1102}
D.~L. Hill and J.~A. Wheeler, {Nuclear Constitution and the Interpretation of
  Fission Phenomena}, {\em Phys. Rev.} {\bf 89}, 1102 (Mar 1953).

\bibitem{PhysRevC.33.2039}
A.~J. Sierk, {Macroscopic model of rotating nuclei}, {\em Phys. Rev. C} {\bf
  33}, 2039 (Jun 1986).

\bibitem{PhysRevC.67.044316}
K.~Pomorski and J.~Dudek, {Nuclear liquid-drop model and surface-curvature
  effects}, {\em Phys. Rev. C} {\bf 67}, p. 044316 (Apr 2003).

\bibitem{Royer20131}
G.~Royer and A.~Subercaze, {Coefficients of different macro–microscopic mass
  formulas from the AME2012 atomic mass evaluation}, {\em Nucl. Phys. A} {\bf
  917}, 1   (2013).

\bibitem{Mavrodiev:2016upv}
S.~C. Mavrodiev and M.~A. Deliyergiyev, {Modification of the Nuclear Landscape
  in the Inverse Problem Framework using the Generalized Bethe-Weizs\"{a}cker
  Mass Formula}  (2016), \textcolor{blue}{[arXiv:nucl-th/1602.06777]}.

\bibitem{PhysRevC.85.024305}
R.~Chen {\em et~al.}, {Single-nucleon potential decomposition of the nuclear
  symmetry energy}, {\em Phys. Rev. C} {\bf 85}, p. 024305 (Feb 2012),
  \textcolor{blue}{[arXiv:nucl-th/1112.2936]}.

\bibitem{Zhang2013234}
Z.~Zhang and L.-W. Chen, {Constraining the symmetry energy at subsaturation
  densities using isotope binding energy difference and neutron skin
  thickness}, {\em Phys. Lett. B} {\bf 726}, 234   (2013),
  \textcolor{blue}{[arXiv:nucl-th/1302.5327]}.

\bibitem{PhysRevC.89.054320}
S.~E. Agbemava {\em et~al.}, {Global performance of covariant energy density
  functionals: Ground state observables of even-even nuclei and the estimate of
  theoretical uncertainties}, {\em Phys. Rev. C} {\bf 89}, p. 054320 (May
  2014), \textcolor{blue}{[arXiv:nucl-th/1404.4901]}.

\bibitem{Wang2014215}
N.~Wang {\em et~al.}, {Surface diffuseness correction in global mass formula},
  {\em Phys. Lett. B} {\bf 734}, 215   (2014),
  \textcolor{blue}{[arXiv:nucl-th/1405.2616]}.

\bibitem{Erler_nature:2012}
J.~Erler {\em et~al.}, {The limits of the nuclear landscape}, {\em Nature} {\bf
  486}, 509  (28 June 2012).

\bibitem{Wang:2014mra}
R.~Wang and L.-W. Chen, {Positioning the neutron drip line and the $r$-process
  paths in the nuclear landscape}, {\em Phys. Rev. C} {\bf 92}, p. 031303
  (2015), \textcolor{blue}{[arXiv:nucl-th/1410.2498]}.

\bibitem{PhysRevLett.16.197}
G.~T. Garvey and I.~Kelson, New nuclidic mass relationship, {\em Phys. Rev.
  Lett.} {\bf 16}, 197 (Jan 1966).

\bibitem{Janecke1988265}
{J. J\"{a}necke and P.J. Masson}, {Mass predictions from the Garvey-Kelson mass
  relations}, {\em Atom.Data Nucl.Data Tabl.} {\bf 39}, 265   (1988).

\bibitem{Liran1976431}
S.~Liran and N.~Zeldes, A semiempirical shell-model formula, {\em Atom.Data
  Nucl.Data Tabl.} {\bf 17}, 431   (1976).

\bibitem{Nayak1999213}
R.~Nayak and L.~Satpathy, {Mass Predictions in the Infinite Nuclear Matter
  model}, {\em Atom.Data Nucl.Data Tabl.} {\bf 73}, 213   (1999).

\bibitem{Koura200047}
H.~Koura {\em et~al.}, {Nuclear mass formula with shell energies calculated by
  a new method}, {\em Nucl. Phys. A} {\bf 674}, 47   (2000).

\bibitem{PhysRevC.52.R23}
J.~Duflo and A.~Zuker, {Microscopic mass formulas}, {\em Phys. Rev. C} {\bf
  52}, R23 (Jul 1995).

\bibitem{PhysRevC.65.037301}
C.~Samanta and S.~Adhikari, {Extension of the Bethe-Weizs\"acker mass formula
  to light nuclei and some new shell closures}, {\em Phys. Rev. C} {\bf 65}, p.
  037301 (Feb 2002).

\bibitem{ADHIKARI:2004}
S.~Adhikari and C.~Samanta, {Systematic Study of Shell Effect Near Drip-lines},
  {\em Int.J.Mod.Phys. E} {\bf 13}, 987  (2004),
  \textcolor{blue}{[arXiv:nucl-th/0408058]}.

\bibitem{BASU:2004}
D.~N. Basu, {Neutron and Proton Drip Lines Using the Modified
  Bethe-Weizs\"{a}cker Mass Formula}, {\em Int.J.Mod.Phys. E} {\bf 13}, 747
  (2004), \textcolor{blue}{[arXiv:nucl-th/0306061]}.

\bibitem{Oganessian201562}
Y.~Oganessian and V.~Utyonkov, {Superheavy nuclei from $^{48}Ca$-induced
  reactions}, {\em Nucl. Phys. A} {\bf 944}, 62   (2015), Special Issue on
  Superheavy Elements.

\bibitem{Popov2001}
V.~S. Popov, Critical charge in quantum electrodynamics, {\em Physics of Atomic
  Nuclei} {\bf 64}, 367  (2001).

\end{thebibliography}

\end{document}